# Academic Storage Cluster

Alexander von Tottleben
Data & Knowledge Engineering
University of Wuppertal
Wuppertal, Germany
alexander@vontottleben.de

Cornelius Ihle
Data & Knowledge Engineering
University of Wuppertal
Wuppertal, Germany
ihle@gipplab.org

Moritz Schubotz
Dept. of Mathematics
FIZ – Karlsruhe
Berlin, Germany
moritz.schubotz@fiz-karlsruhe.de

Bela Gipp
Data & Knowledge Engineering
University of Wuppertal
Wuppertal, Germany
gipp@uni-wuppertal.de

*Abstract*— Decentralized storage is still rarely used in an academic and educational environment, although it offers better availability than conventional systems. It still happens that data is not available at a certain time due to heavy load or maintenance on university servers. A decentralized solution can help keep the data available and distribute the load among several peers. In our experiment, we created a cluster of containers in Docker to evaluate a private IPFS cluster for an academic data store focusing on availability, GET/PUT performance, and storage needs. As sample data, we used PDF files to analyze the data transport in our peer-to-peer network with Wireshark. We found that a bandwidth of at least 100 kbit/s is required for IPFS to function but recommend at least 1000 kbit/s for smooth operation. Also, the hard disk and memory size should be adapted to the data. Other limiting factors such as CPU power and delay in the internet connection did not affect the operation of the IPFS cluster.

## I. INTRODUCTION

In this project, we identify the benefits and shortcomings of recent decentralized content-addressable storage solutions on the example of IPFS and its suitability to store, retrieve, and manage academic documents. For this purpose, we evaluate the read/write performance and chunk distribution inside a private cluster. Instead of downloading the data from a specific server to another client, a peer asks other nearby peers for the information. Similarly, data is provided by others in the network; hence the information should still be retrievable when a single peer is offline or lost its data. This works because the pinned data is replicated at each cluster peer and addressed by its content.

## II. SYSTEM

In our approach, we simulate a cluster of nodes on a single host. This way, we can control network parameters and monitor resource demands in an isolated and controlled setup. Each node runs the IPFS cluster software, with one node set up as a bootstrap node, so peers initially can join the cluster. We used the virtualization software Docker to set up multiple containers to connect them and control variables on the host. As the host operating system, we chose Ubuntu 20.04.1 [1] because it offers a balance between ease of use, availability of tools, performance, and relatively lightweight in terms of resource requirements.

For the containers, we created a Dockerfile, which is based on alpine and supports a packet manager to extend the instance with software for IPFS [2], IPFS-Cluster, monitoring tools, compilers, network restriction and analyzing tools. To configure the restrictions, we used Traffic Control (tc is part of iproute2). Here we specified the bandwidth and an artificial delay. The CPU and memory configuration is specifiable in Docker. The IPFS daemon is started last on each container. We captured the connections and package of the whole cluster in Wireshark on the host.

## III. INTERPLANETARY FILE SYSTEM

### A. Function

In HTTP clients request the data's location to initiate data retrieval. This location information comes in the form of an IP address and the file's path but usually uses a DNS domain to provide a human-readable identifier. Since the data is specified by its location, it is not guaranteed to be the exact requested data we expect.

IPFS, on the other hand, addresses the data according to the content itself (content-addressing) [3]. For this reason, a fingerprint (content hash) must be created for each file. This allows the data to be uniquely described and verified.

To check whether the data is still available, a provider call is executed by the cluster software at regular intervals. Furthermore, duplicates can be avoided by cleverly dividing the files into chunks and using data structures like the Merkle DAG [4]. Each node stores only the data a user has pinned or cached, and the IDs of neighboring participating peers [3]. Additionally, a bootstrap node or mDNS is required to allow IPFS nodes to initially find other peers to form the private cluster swarm. Public IPFS bootstrap nodes would compromise privacy, hence, we use our private bootstrap-node to onboard new participating peers.

### B. Cluster

The cluster consists of multiple IPFS nodes in a swarm configuration as shown in Figure 1.

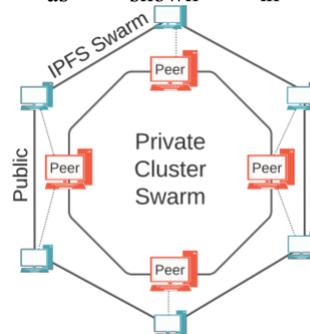

*Figure 1: IPFS swarm and cluster*

Running a cluster like this has the following advantages:
1. A cluster is horizontally scalable by simply adding more peers to distribute the load in the system.
2. Nodes can use a different network connection, location, and power sources to improve the system's robustness.
3. A larger number of replicating cluster peers and the monitoring of the replication improves availability.

*C. Private Cluster*

As default IPFS runs in public mode, so every peer can request blocks with a want-list, and other nodes serve the block if available. Everyone can make data available from their node. Sensitive data however should be shared within a certain group of nodes. In our scenario we want data to be distributed only to selected peers. Also, the environment needs to be consistent so we can change one parameter and measure the impact. However, a private cluster cannot be used by peers other than the ones we initialize. This means that privacy is guaranteed at the price of having the storage space limited to the capacity of the private peers.

## IV. EVALUATION

*A. Methodology*

We measured several system parameters to analyze the IPFS-peers' resource demands. IPFS applies chunking to distribute each file as 256 kB-sized fractions, making the datatype irrelevant. The test file (6.93 MB) was provided by a peer and then pinned by the cluster. We monitored the pinning status until all cluster peers replicated it. Then we deleted the file on a different peer and confirmed it. Finally, we executed a GET from the peer where the file was deleted. For all operations, the duration, quantity of packets, network load, as well as CPU and memory demands were monitored and documented. This way slow networks can be simulated by reducing the bandwidth and increasing the response times artificially. Further, slow peers can be simulated by allocating fewer resources to a particular peer. In Docker changes are easy to make and Wireshark only needs to run on the hosting machine for data investigation. But a simulation has limitations since it cannot mimic the complexity of real peers at different locations.

*B. Results*

By analyzing the traffic in Wireshark (Figure 2), we proved the data distribution to all peers. The traffic data also demonstrated that data was replicated correctly, and the file was divided and transmitted in the form of multiple chunks.

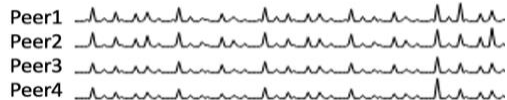

Figure 2: Peer Load Distribution

*1) Bandwidth Limitation:* IPFS worked starting from a bandwidth of about 100 kbit/s, below that the peer was not recognized by others in the network. The more bandwidth was available to a peer, the more likely that peer was to provide chunks in priority to others. This resulted in a larger number of bigger packets. For bandwidths beyond 1 Mbit/s, this effect was no longer noticeable, since the number of transmitted packets from fast (10 Mbit/s) and slow (1 Mbit/s) peers was about the same. During 10 tests with very low bandwidth (approx. 100 kbit/s), we detected that affected peers slow down the entire cluster. Deactivating this peer recovered the download speed. It is advantageous to connect all peers with at least 1 Mbit/s for good response.

*2) Transmission Time Manipulation:* About 20 tests showed that using varying delays (10 to 2000 ms) in the connection to simulate a slow response time did not affect the choice of the peers for data-provisioning. Thus, the peer with a high simulated ping was chosen to provide data just as a peer with a low response time.

*3) System Limitation:* Limiting the CPU did not lead to any measurable difference. However, limiting the available memory had an impact on the system. For example, a running container with a PDF file requires about 200 MB of memory. Since Docker stores most of the running container data in the host machine's memory, bottlenecks can quickly occur with large files. These types of files were therefore not tested, due to the limitations of our test setup. The entire results are available in our GitHub repository [5].

*C. Conclusion*

A cluster can offer benefits over conventional server-client systems like redundancy, easier scalability, and enhanced availability because of its cooperating peers. We created a controlled environment with minimal uncertainties to obtain consistent data. We provide Dockerfiles on GitHub [5] to ensure reproducibility and easy setup in future research. Further, we developed scripts to automate the evaluation and provisioning of peers and tools with a single command. As expected, the bandwidth limitation slowed down download speeds, and the CPU limitation had minimal to no impact. In contrast, we see that the simulated long response time of up to 2 s had no negative effect on the choice of providing peers. Further experiments with real distributed machines are needed to explore whether an IPFS cluster is suitable as a decentralized academic repository.

[1] "FocalFossa/ReleaseNotes/ChangeSummary/20.04.1 - Ubuntu Wiki." https://wiki.ubuntu.com/FocalFossa/ReleaseNotes/ChangeSummary/20.04.1 (accessed Jan. 21, 2021).
[2] "ipfs/go-ipfs - Docker Hub." https://hub.docker.com/r/ipfs/go-ipfs (accessed Jan. 22, 2021).
[3] J. Benet, "IPFS - Content Addressed, Versioned, P2P File System," *ArXiv14073561 Cs*, Jul. 2014, Accessed: Dec. 01, 2020. [Online]. Available: http://arxiv.org/abs/1407.3561
[4] J. Kan and K. S. Kim, "MTFS: Merkle-Tree-Based File System," in *2019 IEEE International Conference on Blockchain and Cryptocurrency (ICBC)*, Seoul, Korea (South), May 2019, pp. 43–47. doi: 10.1109/BLOC.2019.8751389.
[5] "ag-gipp/acst: ACademic-STorage-cluster." https://github.com/ag-gipp/acst (accessed Apr. 28, 2021).